\documentclass[aps,prc,groupedaddress,twocolumn,showpacs,amsmath,amssymb]{revtex4}
\usepackage{graphics}% Include figure files
\usepackage{dcolumn}% Align table columns on decimal point 
\usepackage{longtable}

\begin{document}

\title{Additive empirical parametrization and microscopic study of the deuteron breakup}

\author{M.~Avrigeanu} \email{marilena.avrigeanu@nipne.ro}
\author{V.~Avrigeanu}

\affiliation{Horia Hulubei National Institute for Physics and Nuclear Engineering, P.O. Box MG-6, 077125 Bucharest-Magurele, Romania}

%\date{\today}

\begin{abstract}
\noindent
Comparative assessment of the total breakup proton-emission cross sections measured for 56 MeV deuteron interaction with target nuclei from $^{12}$C to $^{209}$Bi, with an empirical parametrization and recently calculated microscopic neutron-removal cross sections has been done at the same time with similar data measured at 15, 25.5, 70, and 80 MeV.    
Comparable mass dependances of the elastic-breakup (EB) cross sections provided by the empirical parametrization and the microscopic results have been also found at the deuteron energy of 56 MeV, while the assessment of absolute-values variance up to a factor of two has been not possible due to the lack of EB measurements at energies higher than 25.5 MeV. 
While the similarities represent an additional validation of the microscopic calculations, the cross-section difference should be considered within the objectives of further measurements.
\end{abstract}


\maketitle

\section{Introduction}
\label{Sec1}

An update of the theoretical analysis of deuteron-nuclei interaction within an unitary and consistent account of the related reaction mechanisms is highly requested by on-going strategic research programs (ITER, IFMIF, SPIRAL2-NFS)  \cite{iter} and medical investigations using accelerated deuterons, on the basis of improved nuclear databases \cite{FENDL}. The need of this update comes essentially from specific noncompound processes that should be considered in the case of the incident deuterons, making them substantially different from other incident particles. Thus, the deuteron breakup (BU) is particularly quite important due to the large variety of reactions initiated by the breakup nucleons along the whole incident energy range \cite{BU1,BU2,BU3}. Otherwise, the deuteron interaction with low and medium mass target nuclei and incident energies below and around the Coulomb barrier proceeds largely through stripping and pick-up direct reaction (DR) mechanisms, while pre-equilibrium emission (PE) and evaporation from fully equilibrated compound nucleus (CN) become important at higher energies \cite{RC2015}. On the other hand, the scarce deuteron-breakup experimental data systematics \cite{pamp78,wu79,klein81,mats80,must87} related to the high complexity of the breakup mechanism has constrained so far a comprehensive analysis of the deuteron interactions within wide ranges of target nuclei and incident energies.

Moreover, unlike the DR, PE, and CN theoretical models, various current studies concern the theoretical description of the breakup mechanism and its components, namely the elastic breakup (EB), in which the target nucleus stays in its ground state and both deuteron constituents fly apart, and  the inelastic breakup or breakup fusion (BF), where one of these constituents interacts non-elastically with the target nucleus. Microscopic EB calculations have been performed using the continuum-discretized coupled-channels (CDCC) method (\cite{kamimura86,austern,deltuva,ogata} and Refs. therein), treating the deuteron scattering on a target nucleus $A$ by a three-body reaction model. The EB  component is treated as an inelastic excitation of the projectile, due to the nuclear and Coulomb interactions with the target, through the coupled channels approach. In order to deal with a finite set of coupled equations, an essential feature of the CDCC method is the introduction of a discretization procedure, in which the continuum  spectrum is represented by a finite and discrete set of square-integrable functions. However, since the deuteron elastic breakup component is almost one order of magnitude weaker than the total EB+BF process \cite{klein81,must87}, a model of either the total breakup or the inelastic breakup formalism is highly requested. 

The recently deuteron-breakup detailed analyzes of both EB and BF components by the distorted wave Born approximation (DWBA) method \cite{carlson,potel,moro}, with prior/post form amplitudes, performed a successful description of proton spectra and angular distributions for the $(d,p)$ reaction on $^{27}$Al, $^{58}$Ni, $^{93}$Nb, and $^{118}$Sn at incident energies from 15 MeV to 100 MeV. 
The corresponding calculated EB and BF cross sections would be, however, also quite useful for the comparison with experimental data \cite{pamp78,wu79,klein81,mats80}. More recently, Neoh {\it et al.} \cite{neoh} applied the CDCC extension of the eikonal reaction theory (ERT), using also microscopic optical potentials, to the analysis of the EB and neutron removal 
cross sections at 28 MeV/nucleon on various target nuclei from $^{12}$C to $^{209}$Bi, of interest for further studies of unstable nuclei structure. 

A comparative assessment of the measured data and the results of microscopic description of the BU process as well as current parametrization already involved within recent systematic studies of deuteron-induced reactions \cite{BU1,BU2,BU3}, aimed by this work, could be equally useful to basic objectives \cite{neoh} and improved nuclear data calculations within a considerable range of target nuclei and incident energies up to 60 MeV. 
The former parametrization \cite{FED} is addressed in Sec II, including a further normalization of the EB  that has been proved necessary at energies beyond the restricted range of the available measured data, as well as an additional constraint of the total BU cross section for the target nuclei above $A$=200. 
The comparison of the total BU proton-emission cross sections measured by Matsuoka {\it et al.} \cite{mats80}, for 56 MeV deuteron interaction with target nuclei from $^{12}$C to $^{209}$Bi, with the empirical parametrization and the microscopic neutron-removal cross sections \cite{neoh}, done at the same time with similar data measured at 15, 25.5, 70, and 80 MeV is discussed in Sec. III.A, while a similar analysis is given for the EB cross sections in Sec. III.B, followed by conclusions in Sec. IV.

\section{Deuteron breakup parametrization}

\begin{figure} [b]
\resizebox{\columnwidth}{!}{\includegraphics{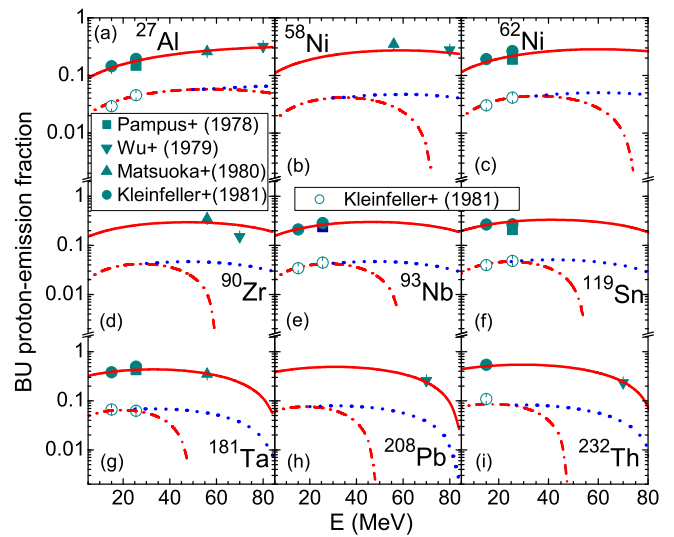}}
\caption{(Color online) Comparison of experimental \cite{pamp78,wu79,mats80,klein81} total breakup proton-emission (solid symbols) and elastic-breakup (open circles) fractions, and the corresponding parametrizations \cite{FED} (solid and dash-dotted curves, respectively) as well as the normalized EB fractions (dotted) for deuterons incident on nuclei from $^{27}$Al (a) to $^{232}$Th (i) at energies up to 80 MeV.}
\label{FED_Ef} 
\end{figure}

An empirical parametrization \cite{FED} of both the total breakup (EB+BF) and EB data has involved the assumption that the inelastic-breakup cross section for neutron emission $\sigma_{BF}^n$ is the same as that for the proton emission $\sigma_{BF}^p$ (e.g., Ref. \cite{must87}), so that the total breakup cross sections $\sigma_{BU}$ is given by the sum $\sigma_{EB}$+2$\sigma_{BF}^{n/p}$. The parametrization has concerned the total BU nucleon-emission and EB fractions, i.e. $f_{BU}^{n/p}$ = $\sigma^{n/p}_{BU}/\sigma_R$ and  $f_{EB}$=$\sigma_{EB}/\sigma_R$, respectively, where $\sigma_{R}$ is the deuteron total-reaction cross section. 
Thus, the dependence of these fractions on the deuteron incident energy $E$ and the target-nucleus atomic $Z$ and mass  $A$ numbers was obtained \cite{FED} through analysis of the experimental systematics of deuteron-induced reactions on target nuclei from $^{27}$Al to $^{232}$Th and incident energies up to 80 MeV for the former \cite{pamp78,wu79,klein81,mats80}, 
\begin{eqnarray}\label{eq:1}
f^{n/p}_{BU}=0.087-0.0066 Z + 0.00163 ZA^{1/3} +  \:  \nonumber \\
0.0017A^{1/3}E - 0.000002 ZE^2 \:\:\:\:  ,
\end{eqnarray}
but within a more restricted energy range up to 30 MeV \cite{klein81,must87} for the later:
\begin{eqnarray}\label{eq:2}
f_{EB}=0.031-0.0028 Z + 0.00051 ZA^{1/3} +  \:  \nonumber \\
0.0005A^{1/3}E - 0.000001 ZE^2 \:\:\:\: . 
\end{eqnarray}
The comparison of the experimental data and parametrization results shown in Fig.~\ref{FED_Ef} for deuterons incident on nuclei from $^{27}$Al to $^{232}$Th, at energies up to 80 MeV, has proven a suitable agreement.

\begin{figure*} %[t]
\resizebox{1.8\columnwidth}{!}{\includegraphics{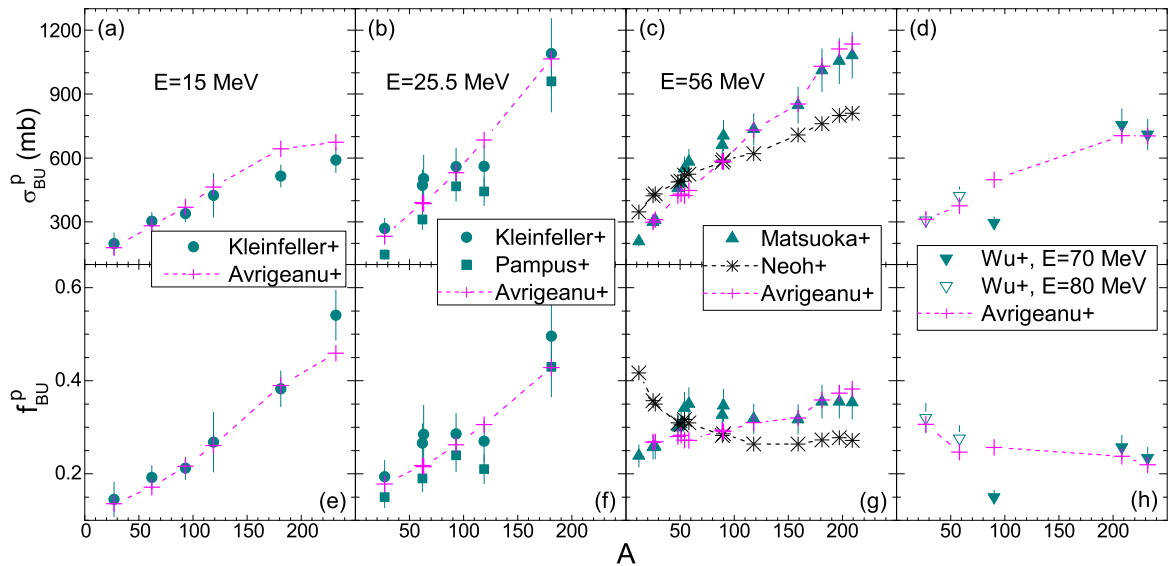}}
\caption{(Color online) Comparison of the mass dependence of the measured \cite{pamp78,wu79,mats80,klein81} total BU proton-emission cross sections (a-d) and fractions (e-h) with the predictions of the microscopic theory for neutron removal cross sections \cite{neoh} (asterisks) and of the empirical parametrization (crosses), connected by dashed lines for eye guiding, for target nuclei from $^{12}$C up to $^{209}$Bi, at the incident energies of 15, 25.5, 56, 70, and 80 MeV.}
\label{Klein_etal_BUpf} 
\end{figure*}

\subsection{Deuteron EB normalization}

On the other hand, it was found an apparent decrease of the fraction $f_{EB}$ at energies beyond the range $E$$<$30 MeV of the EB data \cite{klein81,must87}, unlike the trend of both the fraction $f_{BU}^{p}$ fraction (Fig.~\ref{FED_Ef}) and the total-reaction cross section. Consequently, the correctness of this EB extended parametrization has been checked through the comparison of its predictions and results of the CDCC method for the $^{63}$Cu and $^{93}$Nb target nuclei \cite{maCDCC}. Thus, while a good  agreement was obtained between the EB data \cite{klein81} and both the CDCC results and the empirical parametrization \cite{FED} at the energies of the available data, at higher energies the energy dependence of $\sigma_{R}$  and the $f_{BU}^{p}$ parametrization is common only to the CDCC results \cite{maCDCC}. Therefore, the necessary caution in extrapolating the $f_{EB}$ empirical parametrization beyond the energies of the corresponding data have had to be considered at the same time with the  challenging CDCC calculations for each target/energy of interest. Under these conditions it has been opportune to adopt a normalized EB fraction for the energies beyond the maximum of the former parametrization \cite{FED} by taking into account the energy dependence of the $f_{BU}^{p}$ fraction \cite{ND2016}. Hence, we have chosen to keep unchanged the ratio of the EB and BU fractions at the incident energies above the energy $E_{max}$ corresponding to the maximum of the $f_{EB}$ fraction \cite{FED}, by means of the relation:
\begin{equation}\label{eq:3}
      f_{EB}^{norm}(E) = f_{BU}^{n/p}(E) \frac{f_{EB}(E_{max})}{f_{BU}^{n/p}(E_{max})} , \:\:\:\: E > E_{max} \:\:\:\: .
\end{equation}
%\noindent
%
Thus, the normalized EB fraction follows the behavior of the total BU nucleon-emission fraction shown in Fig.~\ref{FED_Ef}, in agreement with the CDCC calculation results \cite{maCDCC}.  Despite the EB component is less than 10\% of total BU cross section, this $f_{EB}$ normalization is of particular interest at deuteron energies above $\sim$50 MeV and especially for heavier target nuclei, for the inelastic breakup fraction
\begin{equation}\label{eq:4}
 f_{BF}^{n/p}  =  f_{BU}^{n/p}  - f_{EB}^{norm}  \:\:\:\: ,
\end{equation}
%\noindent
%
as well as the total breakup fraction 
\begin{equation}\label{eq:5}
 f_{BU}  =  2 f_{BU}^{n/p} - f_{EB}^{norm}  \:\:\:\: , 
\end{equation}
%\noindent
%
under the above-mentioned assumption of equal neutron- and proton-emission BU cross sections \cite{FED}.

\subsection{Deuteron BU additional constraint}

The so scarce total BU proton-emission systematics for heaviest nuclei ($A$$>$200) at incident energies around the Coulomb barrier, of great interest for deuteron interaction with actinide nuclei \cite{Pad,wilson,ducasse}, includes only one single datum for $^{232}$Th at $E$=15 MeV \cite{klein81}. It is properly described by the present parametrization as well as that at 70 MeV reported by Wu {\it et al.} \cite{wu79} (Fig~\ref{FED_Ef}). However, following the EB fraction normalization, i.e. Eqs. (~\ref{eq:3})  and (\ref{eq:5}), the total BU fraction corresponding to this target nucleus  exceeds unity (e.g.,  $f_{BU}$=1.0215 at $E$=32 MeV). 

This unphysical overrun of the total-reaction cross sections should be firstly considered with respect to the systematics accuracy. Since the corresponding data errors amount to 10-15\% \cite{FED}, we have adopted an additional constraint for $A$$>$200, namely that the $f_{BU}$ fraction should not exceed 0.9. This figure is only presently a simple hard limit which has been also included in the latest version of the computer code TALYS$-$1.8 \cite{talys} as well as the above-mentioned EB normalization. Nevertheless, it should be confirmed by further data measurements and also theoretical modeling progress.

\section{Microscopic and parametrization comparison}
\subsection{The total BU proton emission}

The comparison of the total BU proton-emission cross sections $\sigma^p_{BU}$ measured by Matsuoka {\it et al.} \cite{mats80}, for 56 MeV deuteron interaction with target nuclei from $^{12}$C to $^{209}$Bi, with the above-described parametrization and the microscopic neutron-removal cross sections \cite{neoh} is shown in Fig.~\ref{Klein_etal_BUpf} (a-d) at the same time with similar data measured at 15 \cite{klein81}, 25.5 \cite{pamp78,klein81}, and 70 and 80 MeV \cite{wu79}.   
Since the absolute cross sections may depend on the model ingredients of reaction mechanisms involved within the experimental data analysis, e.g., optical and PE model parameters, a similar comparative analysis concerns at the same time in Fig.~\ref{Klein_etal_BUpf} (e-h) the corresponding total BU proton-emission fractions $f_{BU}^{p}$. 
On the other hand, the $f_{BU}^{p}$ values may illustrate the importance of the breakup process among the other reaction mechanisms related to the deuteron interaction. 
Moreover, the same scale is used for the $\sigma^p_{BU}$ as well as $f_{BU}^{p}$ values at all incident energies of the available experimental data, in order to make possible also an assessment of their energy dependence.

\begin{figure} %[b]
\resizebox{\columnwidth}{!}{\includegraphics{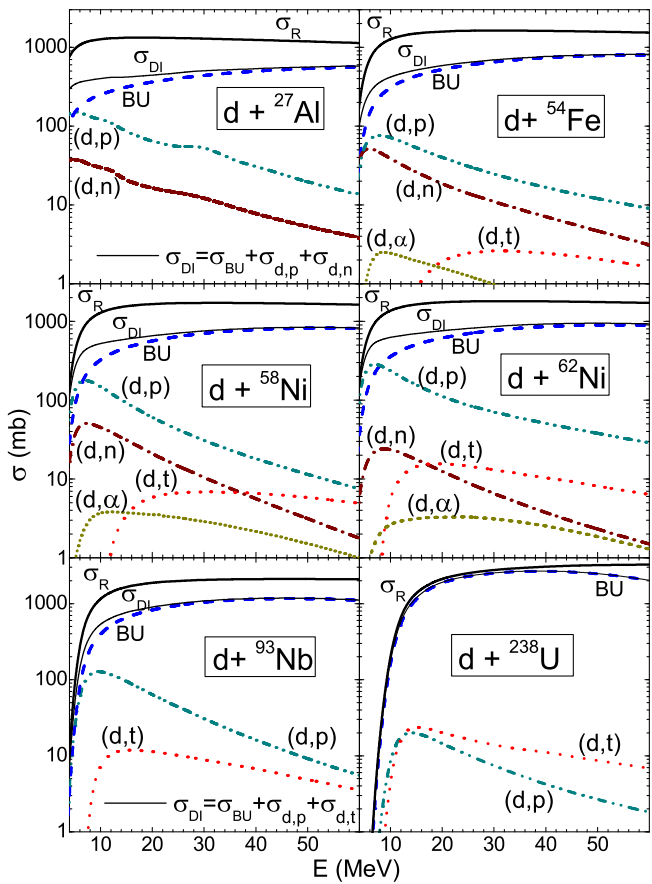}}
\caption{(Color online) Total-reaction (thick solid curves), direct interactions (DI, thin-solid curve), BU (dashed curves), stripping $(d,p)$ (dash-dot-dotted curves) and $(d,n)$ (dash-dotted curves), and pick-up $(d,t)$ (dotted curves) and $(d,\alpha)$ (short-dotted curves) reaction cross sections for deuterons on $^{27}$Al, $^{54}$Fe, $^{58}$Ni, $^{64}$Ni, $^{93}$Nb, and $^{238}$U \cite{BU1,BU2,BU3,ND2016} (see text).}
\label{Ald_Fed_Nid_Nbd_Ud} 
\end{figure}

There are several features which are pointed out by this comparative analysis.
First, the increase of $\sigma^p_{BU}$ with the mass of the target nucleus is well described by the empirical parametrization for all deuteron energies from 15 to 80 MeV. There is a similar trend of the microscopic results for medium-mass nuclei with  $40<$$A<120$, while it is apparent an overestimation of the measured data for light nuclei ($A$$<$40) as well as an underestimation for heavier ones ($A$$>$120).

Second, the $f_{BU}^{p}$ values show that the importance of the BU mechanism 
is increasing with the target-nucleus mass, from $^{27}$Al up to $^{232}$Th, at the lower incident energies of 15 and 25.5 MeV. This increase is less significant at the energy of 56 MeV, and even reversed at 70-80 MeV. Actually it seems that the fraction $f_{BU}^{p}$ has reached its maximum   at 56 MeV, for the target nuclei with $A$$>$120, while for 40$<$$A$$<$120 this maximum moves at energies over 56 MeV but lower than 70-80 MeV. 
Moreover, the $f_{BU}^{p}$ values are still increasing with the incident energy even at 80 MeV for the deuteron interaction with light target nuclei ($A$$<$40).
These energy dependences of the measured $f_{BU}^{p}$, which are obvious also in Fig.~\ref{FED_Ef} for the target nuclei from $^{27}$Al up to $^{232}$Th, are satisfactorily described by the empirical parametrization. The microscopic results at the energy of 56 MeV \cite{neoh}  show a steep decrease for target nuclei from $A$=12 up to $A$$\sim$120, apart from the data, while for $A$$>$120 their underestimated values describe however the target-nucleus mass dependence. Thus, one may note that the microscopic theory provides at the incident energy of 56 MeV a mass dependence which becomes real at the higher energies of 70-80 MeV.

A comment should concern the inclusion of the $(d,p)$ stripping direct reaction by the  microscopic total BU proton-emission cross sections, unlike the experimental data \cite{pamp78,wu79,mats80,klein81} which were obtained through a distinct analysis of the BU, DR, PE, and CN reaction mechanisms. However, the results of systematic analysis of the four different mechanisms for the deuteron interactions with $^{27}$Al, $^{54,56-58}$Fe, $^{58,60-62,64}$Ni, $^{63,65}$Cu, $^{93}$Nb,  and $^{238}$U \cite{BU1,BU2,BU3,ND2016} (e.g., Fig.~\ref{Ald_Fed_Nid_Nbd_Ud}) point out the decreasing importance of the stripping reaction with the deuteron energy increase. Thus, the corrections for $(d,p)$ DR contribution should be less than 1\% at the energy of $56$ MeV, with no real effect on the comparison shown in Fig.~\ref{Klein_etal_BUpf}.

\subsection{The elastic breakup}

\begin{figure} %[b]
\resizebox{0.65\columnwidth}{!}{\includegraphics{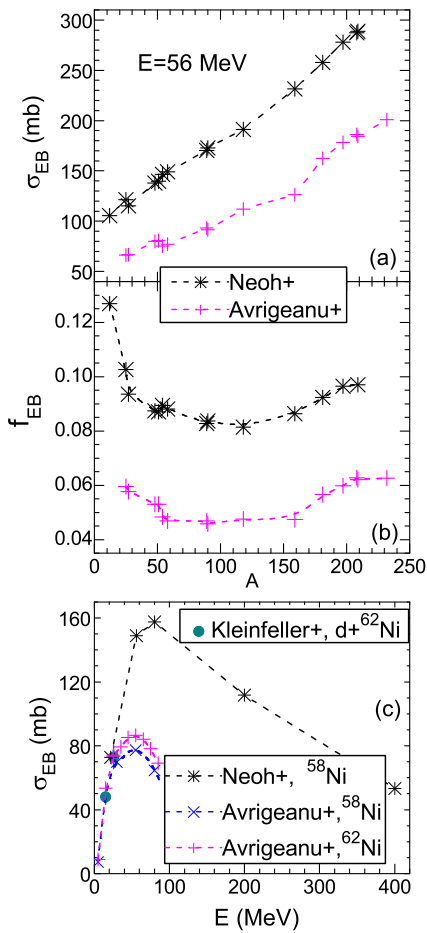}}
\caption{(Color online) Comparison of the mass dependence of the EB cross sections (a) and fractions (b) predicted by the microscopic theory \cite{neoh} (asterisks) and empirical parametrization (crosses), connected by dashed lines for eye guiding, for deuterons incident on target nuclei from $^{12}$C up to $^{232}$Th, at the energy of 56 MeV, as well as of (c) the excitation functions of the EB cross sections measured for deuterons on $^{62}$Ni \cite{klein81}, and the microscopic \cite{neoh} and phenomenological predictions for $^{58,62}$Ni target nuclei (see text).}
\label{EB} 
\end{figure}

A comparison of the mass dependance of the EB cross sections and corresponding fractions $f_{EB}$ provided by the above-described parametrization and the microscopic results \cite{neoh} is shown in Fig.~\ref{EB}(a,b) for the deuteron energy of 56 MeV. Comparable trends of these results are obvious, the theoretical values being larger up to a factor of two. 
Moreover, there have been obtained similarly larger fractions $f_{EB}$ for light ($A$$<$50) and heavier ($A$$>$160) target nuclei. 
Unfortunately, the lack of EB measurements at energies higher than 25.5 MeV does make difficult the assessment of the apparent discrepancy among the microscopic and parametrization predictions for EB cross sections corresponding to the energy of 56 MeV. On the other hand, the parametrized predictions have already been involved within systematic analysis of all available data for deuteron interaction with various nuclei \cite{BU1,BU2,BU3,ND2016}, with a general good agreement between the measured and calculated data. Thus, the just above-mentioned similarities represent an additional validation of the microscopic calculations, while the absolute-value variance should be considered within the objectives of further measurements.

Furthermore, the comparison in Fig.~\ref{EB}(c) of the microscopic \cite{neoh} and empirical EB excitation functions corresponding to the deuteron interaction with $^{58}$Ni nucleus proves the maximum pointed out within the discussion of the total BU protons-emission component. However, the comparison of the EB measurements of Kleinfeller {\it et al.} \cite{klein81} for $^{62}$Ni at the incident energies of only 15 and 25.5 MeV, and the related empirical excitation function which describes these two data points, do not allow a certain assertion concerning neither the deuteron energy corresponding to this maximum, nor the accuracy of the theoretical or empirical predictions.

\section{Conclusions}

Comparative assessment of the measured data and the results of microscopic description \cite{neoh} of the deuteron BU process as well as current parametrization  \cite{FED,ND2016} already involved within recent systematic studies of deuteron-induced reactions \cite{BU1,BU2,BU3} has been carried on. 
A normalization of the EB parametrization \cite{FED} has been proved necessary at energies beyond the restricted range of the available measured data. On the other hand, an additional constraint has to concern the total BU cross section for the target nuclei above $A$=200, at 0.9 of the total-reaction cross section.

The comparison of the total BU proton-emission cross sections $\sigma^p_{BU}$ measured by Matsuoka {\it et al.} \cite{mats80}, for 56 MeV deuteron interaction with target nuclei from $^{12}$C to $^{209}$Bi, with the empirical parametrization and the microscopic neutron-removal cross sections \cite{neoh} has been done at the same time with similar data measured at 15 \cite{klein81}, 25.5 \cite{pamp78,klein81}, and 70 and 80 MeV \cite{wu79}.   
Actually, the total BU proton-emission cross sections measured by Matsuoka {\it et al.} \cite{mats80} at 56 MeV have been essential for the evidence of the maximum of the deuteron breakup mechanism around this incident energy and medium-mass target nuclei. Moreover, the opportunity of the comparison of microscopic and parametrization results at this incident energy has been most useful for further development of both methods of deuteron-breakup mechanism study. At the same time, the corrections for $(d,p)$ stripping-reaction contribution to the neutron-removal cross sections have been shown to be less than 1\% at the energy of $56$ MeV, with no real effect on the above-mentioned comparison.

Comparable mass dependances of the EB cross sections provided by the empirical parametrization and the microscopic results \cite{neoh} have been found at the deuteron energy of 56 MeV, while the assessment of a variance of the absolute values up to a factor of two has been not possible due to the lack of EB measurements at energies higher than 25.5 MeV. However, since the parametrized predictions have already been involved within successful analysis of all available data for deuteron interaction with various nuclei \cite{BU1,BU2,BU3,ND2016}, these similarities represent an additional validation of the microscopic calculations, while the cross-section difference should be considered within the objectives of further measurements.

\section{Acknowledgments}

The authors acknowledge the useful correspondence with Constance Kalbach on need of Sec. II.A. This work has been partly supported by Autoritatea Nationala pentru Cercetare Stiintifica (PN-16420102) and partly carried out within the framework of the EUROfusion Consortium and has received funding from the Euratom research and training programme 2014-2018 under grant agreement No 633053. The views and opinions expressed herein do not necessarily reflect those of the European Commission.

\end{document}